\documentclass[pre,twocolumn,aps,superscriptaddress]{revtex4}

\usepackage{graphicx}
\usepackage{amssymb,latexsym,amsthm,amsmath}    
\usepackage{color}

\begin{document}

\title{Explosive synchronization in weighted complex networks}
\author{I. Leyva}
\affiliation{Complex Systems Group, Univ. Rey Juan Carlos, 28933 M\'ostoles, Madrid, Spain}
\affiliation{Center for Biomedical Technology, Univ. Polit\'ecnica de Madrid, 28223 Pozuelo de Alarc\'on, Madrid, Spain}
\author{I. Sendi\~na-Nadal}
\affiliation{Complex Systems Group, Univ.  Rey Juan Carlos, 28933 M\'ostoles, Madrid, Spain}
\affiliation{Center for Biomedical Technology, Univ. Polit\'ecnica de Madrid, 28223 Pozuelo de Alarc\'on, Madrid, Spain}
\author{J. A. Almendral}
\affiliation{Complex Systems Group, Univ.  Rey Juan Carlos, 28933 M\'ostoles, Madrid, Spain}
\affiliation{Center for Biomedical Technology, Univ. Polit\'ecnica de Madrid, 28223 Pozuelo de Alarc\'on, Madrid, Spain}
\author{A. Navas}
\affiliation{Center for Biomedical Technology, Univ. Polit\'ecnica de Madrid, 28223 Pozuelo de Alarc\'on, Madrid, Spain}
\author{S. Olmi}
\affiliation{CNR-Institute of Complex Systems, Via Madonna del Piano,
  10, 50019 Sesto Fiorentino, Florence, Italy}
\author{S. Boccaletti} 
\affiliation{CNR-Institute of Complex Systems, Via Madonna del Piano, 10, 50019 Sesto Fiorentino, Florence, Italy}

\begin{abstract}
The emergence of dynamical abrupt transitions in the macroscopic state of a system is currently a subject of the utmost interest. Given a set of phase oscillators networking with a generic wiring of connections and displaying a generic frequency distribution, we show how combining dynamical local information on frequency mismatches and global information on the graph topology suggests a judicious and yet practical weighting procedure which is able to induce and enhance explosive, irreversible, transitions to synchronization. We report extensive numerical and analytical evidence of the validity and scalability of such a procedure for different initial frequency distributions, for both homogeneous and heterogeneous networks, as well as for both linear and non linear weighting functions. We furthermore report on the possibility of parametrically controlling the width and extent of the hysteretic region of coexistence of the unsynchronized and synchronized states.

PACS: 89.75.Hc, 89.75.Kd, 89.75.Da, 64.60.an,05.45.Xt.
\end{abstract}

\maketitle

\section{Introduction}

One of the most significant challenges of present-day research is
bringing to light the processes underlying the spontaneous organization of networked dynamical units.
When a network passes from one to another collective {\it phase} under the action of a control parameter, the nature of the associated phase transition is disclosed by the behavior of the order parameter at criticality: continuous for second-order transitions and discontinuous for first-order ones.  In complex networks' theory \cite{Boccaletti2006,Dorogovtsev2008} such phase transitions have been observed in the way a graph collectively organizes its architecture through percolation \cite{Karsai2007,Achlioptas09,various2}, and its dynamical state through synchronization \cite{Boccaletti2002,Arenas2008}.

Abrupt transitions to synchronized states of networked phase oscillators were initially reported in a Kuramoto model \cite{kuramoto} for a particular realization of a uniform frequency distribution (evenly spaced frequencies) and an all-to-all network topology \cite{pazo}. Later on, the same finding was also described for both periodic \cite{jesus1} and chaotic  \cite{leyva12} phase oscillators in the yet particular condition of a heterogeneous degree-distribution with positive correlations between the node degree and the corresponding oscillator's natural frequency.
Recently, Ref. \cite{Zhang13} introduced a more general framework where explosive synchronization (ES) is obtained in weighted networks, where weights are selected to be proportional to the absolute value of the frequency of the oscillators in a way that produces positive correlations between the node strength and the frequency of the oscillator.
\begin{figure}[ht!]
  \includegraphics[width=0.4\textwidth]{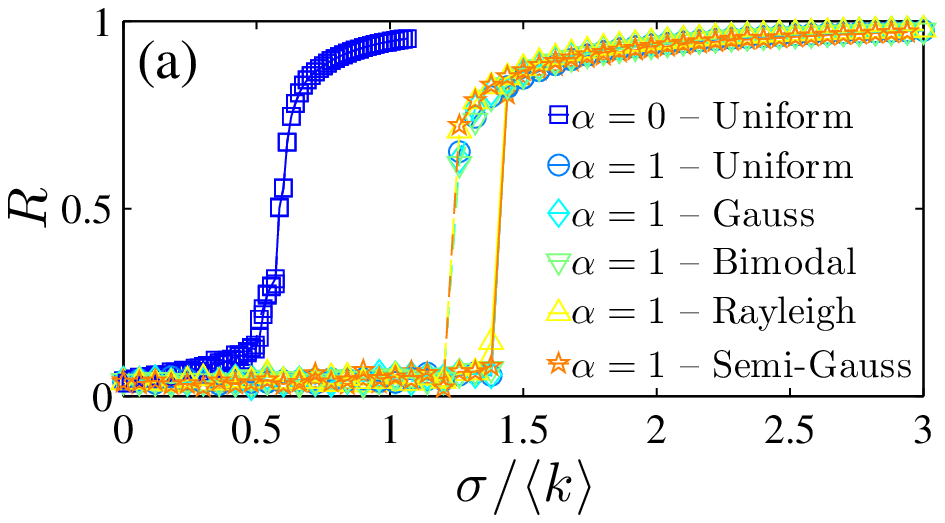}
  \includegraphics[width=0.4\textwidth]{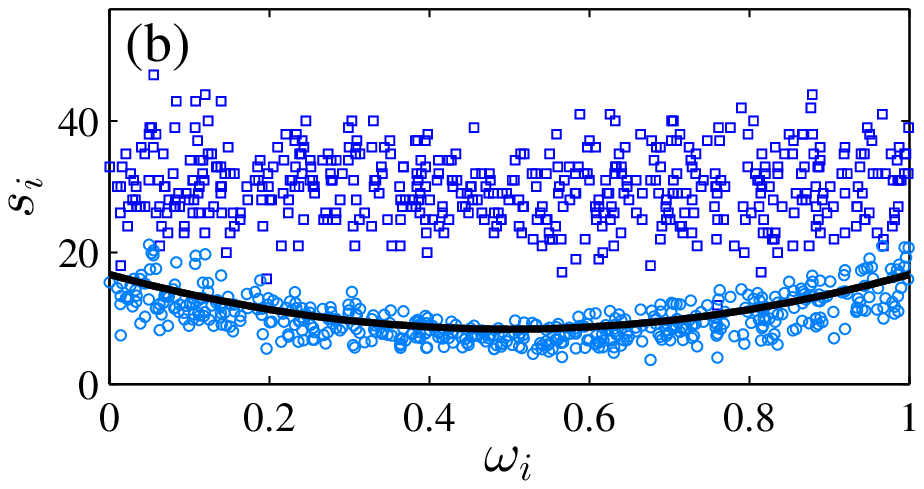}
  \caption{(Color online). (a) Synchronization transitions for N=500
    ER networks, $\langle k\rangle $=30, for un-weighted case ($\alpha=0$)
    (blue squares), and linearly weighted cases ($\alpha=1$) with several frequency
distributions within the range $[0,1]$: uniform, Gaussian,
Gaussian-derived, Rayleigh and semi-Gaussian. Solid and  dashed lines refer to
the forward and backward simulations, respectively. (b) Node strengths
$s_i$ (see text for definition) vs. natural frequencies $\omega_i$,
for the un-weighted (dark blue squares) and weighted (light bue circles) networks
reported in (a). Solid line is proportional to the analytical
prediction $(\omega-\frac{a}{2})^2+\frac{1}{4a}$ in the
thermodynamical limit of our model, with $a=1$ the width of the uniform frequency distribution (see text for more details).}
\label{fig:intro}
 \end{figure}

The weighting procedure proposed in Ref. \cite{Zhang13} inherently
asymmetrizes each link of the network, favoring the interaction
directions from higher to lower frequencies.
In this work, we propose an alternative general framework for
ES in complex networks, based on a weighting
procedure which instead keeps the symmetric nature of the links. The
method is inspired by our recent study of Ref. \cite{leyvanat}, where
it is shown that ES can be obtained
for any given frequency distribution, provided the connection network
is constructed following a rule of {\it frequency disassortativity},
that is, that the synchronization clustering formation is prevented
avoiding close frequencies to couple, in a network generation scheme
ruled by dynamical properties, as the Achlioptas rule
\cite{Achlioptas09} works for the structural case in explosive percolation.

We here deal with the more general case of a network with given
frequency distribution and architecture, and we show that a weighting
procedure on the existing links, that combines information on the
frequency mismatch of the two end oscillators of a link with that of
the link betweenness, has the effect of inducing or enhancing
ES phenomena for both homogeneous and
heterogeneous graph topologies, as well as any  symmetric or
asymmetric frequency distribution. In addition, we show the general scaling properties of the obtained transition, and provide analytical arguments in support of our claims.

\section{Model and numerical results}

 Without lack of generality, our reference is a network $\cal{G}$ of $N$ Kuramoto \cite{kuramoto} phase oscillators, described by:
\begin{equation}
  \frac{d\theta_i}{dt}=\omega_i + \frac{\sigma}{\left< k \right>}\sum_{i=1}^N \Omega_{ij}^{\alpha} \sin(\theta_j-\theta_i),
  \label{eq:kuramoto}
\end{equation}
where $\theta_i$ is the phase of the $i^{th}$ oscillator ($i=1,...,N$),
$\omega_i$ is its associated natural frequency drawn from a frequency
distribution $g(\omega)$, $\sigma$
is the coupling strength, $\langle k\rangle $ is the graph average connectivity ($\langle k\rangle  \equiv \frac{2L}{N}$, with $L$ being the total number of links), and 
\begin{equation}
\Omega_{ij}^{\alpha}=a_{ij}|\omega_i-\omega_j|^{\alpha},
\label{eq:weight1}
\end{equation}
is the weighted link for nodes $i,j$,  being $a_{ij}$ the elements of the adjacency matrix that uniquely defines $\cal{G}$ and 
$\alpha$ a constant parameter which eventually modulates the weight.  The strength of the $i^{th}$
node (the sum of all its links weights) is then
$s_i=\sum_{j} \Omega_{ij}^{\alpha}$.
The classical order parameter for system (\ref{eq:kuramoto}) is $r(t)=
\frac{1}{N} |\sum_{j=1}^N e^{i \theta_j(t)} |$, and the level
of synchronization can be monitored by looking at the value of
$R=\langle r(t)\rangle_T$, with $\langle ...\rangle _T$ denoting a time average over a
conveniently large time span $T$.

As the coupling strength $\sigma$ increases, system (\ref{eq:kuramoto})
undergoes a phase transition at a critical value $\sigma_c$ from the
unsynchronized ($R \sim
1/\sqrt{N}$) to the synchronous ($R = 1$) state, where all oscillators ultimately acquire the same frequency.
In the following, we will describe the nature of such a transition as a function of the re-scaled order parameter $\sigma/\langle k\rangle$.
As for the stipulations followed in our simulations, the state of the
network is monitored by gradually increasing $\sigma$ in steps $\delta
\sigma=0.0005$, starting at $\sigma=0$. Whenever a step $\delta
\sigma$ is made, a long transient (200 time units) is discarded before
the data are recorded and processed.
Moreover, as we are focusing on abrupt, irreversible transitions (and
thus on expected associated hysteretic phenomena),
we perform the simulations also in the reverse way, i.e. starting from a given value $\sigma_{\max}$ (where $R = 1$), and gradually decreasing the coupling by $\delta \sigma$ at each step. In what follows, the two sets of numerical trials are termed as {\em forward} and {\em backward}, respectively.

\subsection{Homogeneous networks}

\begin{figure}
\includegraphics[width=0.4\textwidth]{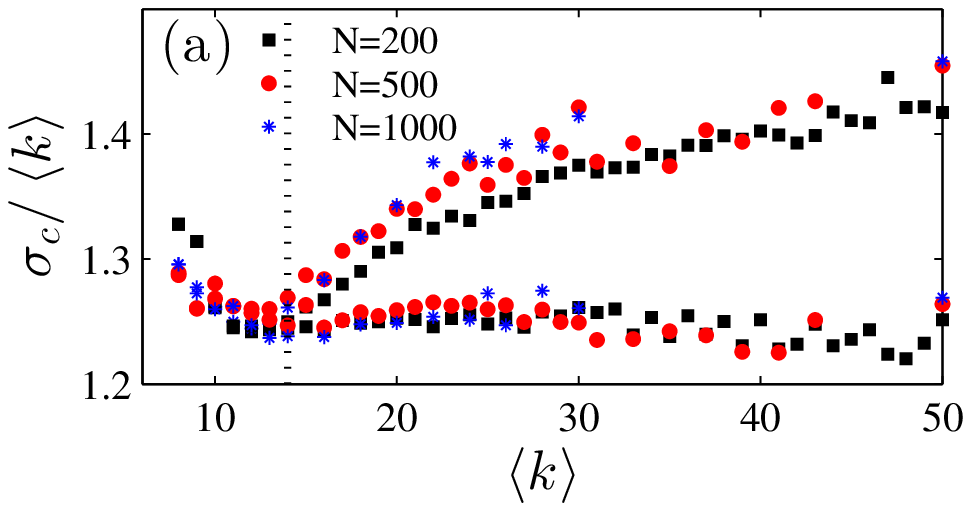}
\includegraphics[width=0.4\textwidth]{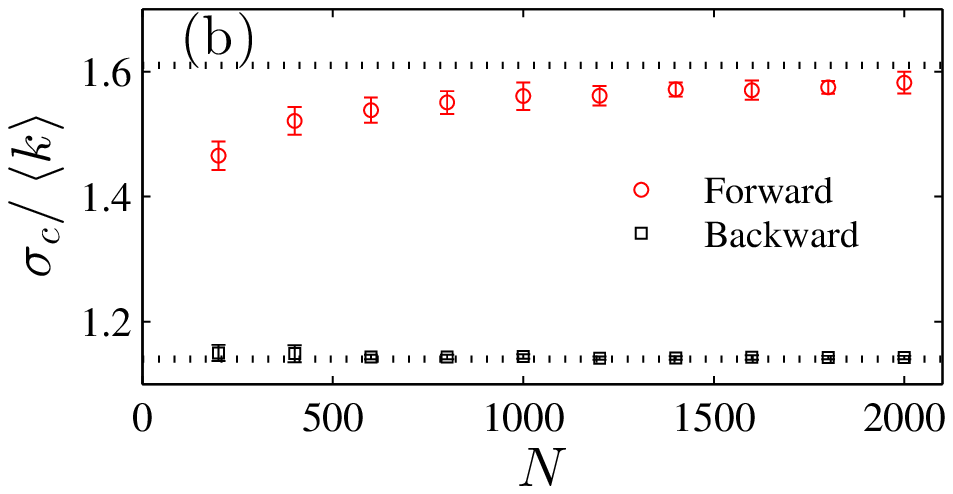}
\caption{(Color online) Critical scaled coupling
    $\sigma_c/\langle k\rangle$ at the onset of
    synchronization/desynchronization using a linear weighting procedure
    $\Omega_{ij}$ ($\alpha=1$) as a function of (a) $\langle
    k\rangle$ for several ER network sizes $N$, and of (b)  $N$ in
    all-to-all coupled networks. In (a) vertical dashed line
marks the passage from a smooth to an explosive phase transition. Both
in (a) and (b) upper and lower branches correspond to {\it forward}
and {\it backward} simulations, respectively.  Each dot accounts for
an average of at least 20 independent runs of uniform frequency
distributions. Horizontal dashed lines in (b) are close to the
analytical values defining the range of the hysteresis in the thermodynamical limit for the Kuramoto model (see explanation in the text). Frequencies are uniformly distributed in the
range $[0,1]$.
\label{fig:scaling}}
\end{figure}

We first report our results on the case of homogeneous graph
topologies. For this purpose, we consider Erd\"os-R\'enyi (ER)
random networks \cite{erdos} of size $N$, and we
describe how an explosive transition is induced,
for sufficiently large values
of $\langle k\rangle $ and irrespectively on the specific frequency distribution
$g(\omega)$. Figure~\ref{fig:intro}(a) reports the results for $N=500$ and several
 frequency distributions $g(\omega)$ within the range $[0,1]$.
For the simplest case of uniform frequency distribution $g(\omega)$=1, while the un-weighted network ($\alpha=0$ in Eq. \ref{eq:weight1}) displays a
smooth, second-order like transition to synchronization [dark blue curve in
Fig. \ref{fig:intro}(a)], the effect of a linear weighting
($\alpha=1$)  is that of inducing a sharp transition in the system,
with an associated hysteresis in the forward
(solid line) and backward (dashed line) simulations.
This drastic change in the nature of the transition is independent of the
frequency distribution $g(\omega)$, as long as they are defined in the
same frequency range $[0,1]$ as shown in Fig.~\ref{fig:intro}(a). The results are identical for symmetric distributions
(homogeneous, Gaussian, a bimodal distribution derived from a Gaussian) and for asymmetric frequency distributions (Rayleigh, a Gaussian centered at $0$ but just using the positive half). See details of the used frequency distributions in \cite{distributions}.

Figure \ref{fig:intro}(b) accounts for the existence of a parabolic
relationship between the strengths and the natural
frequencies of the oscillators associated with the passage from a
smooth to an explosive phase transition. This relationship has been
obtained analytically (see Eq.(\ref{fitstrength})) in the
thermodynamical limit of the Kuramoto model and perfectly fits the
numerical results shown as a solid line in Fig.~\ref{fig:intro}(b). 
It has to be remarked that, while in Ref. \cite{jesus1}
degree-frequency correlation features were imposed to determine
explosiveness in the transition to synchronization,
here the effect of the weighting is to let these topological/dynamical
correlation features {\it spontaneously} emerge,
with the result of shaping a bipartite-like network where low and high frequency oscillators are the ones with maximal overall strength.

Further information about the nature and scaling properties of the transition induced by the linear weighting procedure is gained from Fig.~\ref{fig:scaling}, where it is shown the dependence of  the  scaled critical
coupling $\sigma_c/\langle k\rangle$ on the average connectivity
$\langle k\rangle $ and on the network size $N$. Precisely, Fig.~\ref{fig:scaling}(a) shows that, independently on $N$,
a dynamical bifurcation exists at ${\langle k\rangle } \sim
17$, corresponding to the passage from a second- to a first-order like phase transition. For the
latter regime, the two branches expanding from
 $\langle k \rangle \gtrsim
17$ are associated to the hysteresis in the forward and backward
simulations. The relative independence on $N$ can be explained considering that an important condition for ES to occur is that each node neighborhood must represent a statistically significant sample of the network frequencies up to give a close enough approximation to the global mean frequency, and therefore the synchronization frequency. To reach this target, the required sampling size $n$ for a given population size $N$  is usually calculated with the following formula \cite{Cochran77}
\begin{equation*}
n = \frac{N}{1+ C^2 (N-1)},
\end{equation*}
where $C:=2e/z_{\alpha/2}$, being $e$ the error allowed, $1-\alpha$
the confidence level, and $z_{\alpha/2}$ the upper $\alpha/2$
percentage point of the standard normal distribution. Aside from the
technical details, the important feature in the expression is that the
sampling size converges to a finite value, even for an infinite
population. This is exactly what Fig.~\ref{fig:scaling}(a) shows. Once
the mean degree is large enough, each node has a neighborhood assuring
that its neighbor frequency average is statistically
accurate. Precisely, Fig.~\ref{fig:scaling}(a) suggests that $C \approx
0.24$, indicating that, for mean degrees greater than $\sim$17, each
node has a sufficiently large neighborhood \emph{independently} of the
population size $N$. Figure~\ref{fig:scaling}(b) shows how the scaled critical
couplings defining the hysteresis of the ES transition converge to constant values for the
Kuramoto model (all to all coupling) when $N$ increases which are
quite close to those obtained in the thermodynamical limit of the
Kuramoto model discussed in the analytical section.

\begin{figure}
 \includegraphics[width=0.4\textwidth]{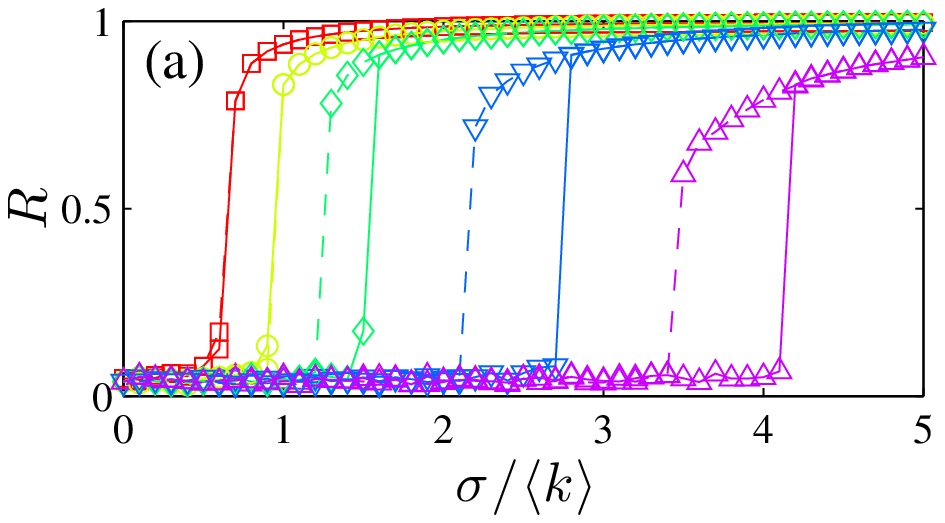}
 \includegraphics[width=0.4\textwidth]{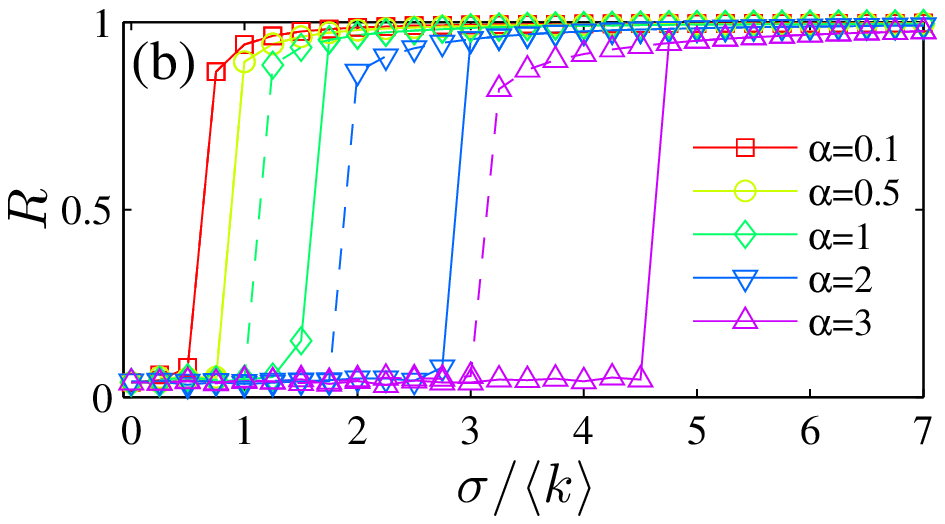}
    \caption{(Color online). Synchronization transitions for ER networks,
      $N=500$, uniformly distributed frequencies in the
      $[0,1]$ range, and nonlinear weighting functions
      $\Omega_{ij}^{\alpha}$. Both plots consider several $\alpha$
      values, from sub-linear to super-linear weighting (see legend in
      panel b). (a) ER networks, $\langle k\rangle =30$, (b) regular
      random networks, $k=30$. In all cases, forward and backward simulations correspond respectively to solid and dashed lines.}
\label{fig:alpha}
 \end{figure}

Furthermore, the weighting procedure inducing ES
is quite general, as a large family of detuning dependent functions
can be used. As an example, Fig.~\ref{fig:alpha} describes the case
of nonlinear weighting procedures, that is, $\alpha\neq 1$ in Eq. (\ref{eq:weight1}). There, we set again $N=500$ and
$\langle k\rangle =30$ and consider
both ER graphs (Fig. \ref{fig:alpha}(a)), and a regular random network  (Fig.~\ref{fig:alpha}(b)), i.e. a network where
each node has exactly the same number  of connections ($k_i=\langle
k\rangle =30$) with the rest of the graph. This latter case has been
obtained by a simple {\it configuration model} \cite{configuration},
imposing a $\delta$-Dirac degree distribution. 
The results in Fig.~\ref{fig:alpha} show that the generic non-linear
function of the frequency mismatch  
given by Eq.~(\ref{eq:weight1}) is able to induce ES in both topologies, and that the effect of a super-linear ($\alpha>1$) weighting (a sub-linear ($\alpha<1$) weighting) is that of enhancing (reducing) the width of the hysteretic region.

\subsection{Heterogeneous networks}
So far, we have considered only homogeneous degree distributions. In order to properly describe the passage from a homogeneous to a heterogeneous degree distribution, we rely
on the procedure introduced in Ref. \cite{Gomez-Gardenes06}. Such a technique, indeed, allows constructing graphs with the same
average connectivity $\langle k\rangle$, and grants one the option of continuously interpolating from
ER to scale-free (SF) networks \cite{Albert1999}, by tuning a single parameter $0 \leq p \leq 1$.
With this method, networks are grown
from an initial small clique, by sequentially adding nodes, up to the desired graph size. Each newly added node
has a probability $p$ of forming random connections with already existing vertices, and a probability
$1-p$ of following a {\it preferential attachment} rule \cite{Albert1999} for the selection of
its connections. As a result, the limit $p=1$ induces an ER configuration, whereas the limit $p=0$ corresponds to a SF network with
degree distribution $P(k) \sim k^{-3}$.
\begin{figure}
  \centering
  \includegraphics[width=0.4\textwidth]{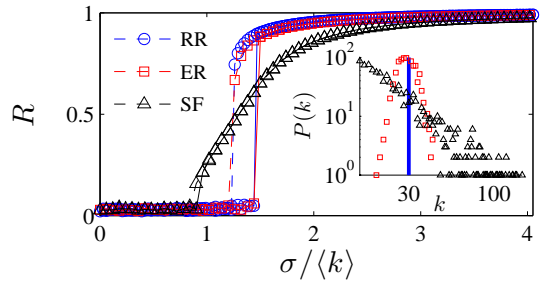}
\caption[]{(Color online) Explosive synchronization  {\it vs.} degree
  heterogeneity. Synchronization transitions as a function of the
  coupling strength for linearly weighted networks ($\alpha=1$) with the same average connectivity
  $\langle k\rangle=30$, but a different second moment of the degree
  distribution:  a regular random (RR) network  with homogeneous degree $\sigma_k=0$ (blue circles), an ER
  network (red squares) and a SF (black triangles). In all cases,
  forward and backward simulations correspond respectively to
  solid and dashed lines. Inset: log-log plot of  the three corresponding degree distributions.
 \label{fig:hetero}}
\end{figure}

Let us set $N=1000$ and $\langle k\rangle =30$ and, after the network construction, let us randomly distribute the oscillators' frequencies in the interval $[0,1]$ and use again a linear weighting function $\Omega_{ij}=a_{ij}|\omega_i-\omega_j|$.
The comparative results are reported in Fig.~\ref{fig:hetero}, from
which it is easy to see that heterogeneity in the degree distribution
actually opposes the onset of explosive synchronization. A similar qualitative scenario (not shown) is obtained
also for different frequency distributions, network's sizes, and (super-linear or sub-linear) weighting functions, allowing one to conclude that heterogeneous degree-distributions require a different weighting approach, where the information on frequency mismatch has to be properly combined with local or global information on the network topology.

\begin{figure}
  \centering
{\includegraphics[width=0.4\textwidth]{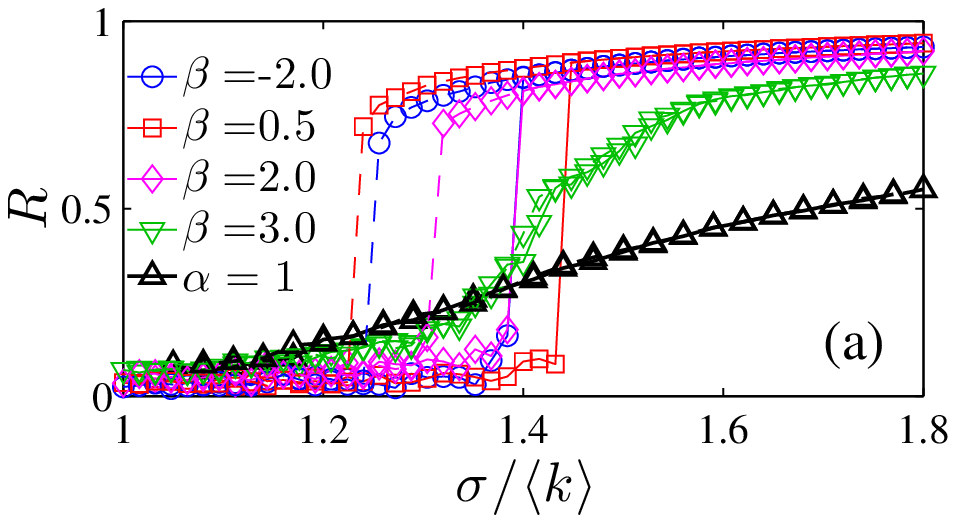}}
 \includegraphics[width=0.4\textwidth]{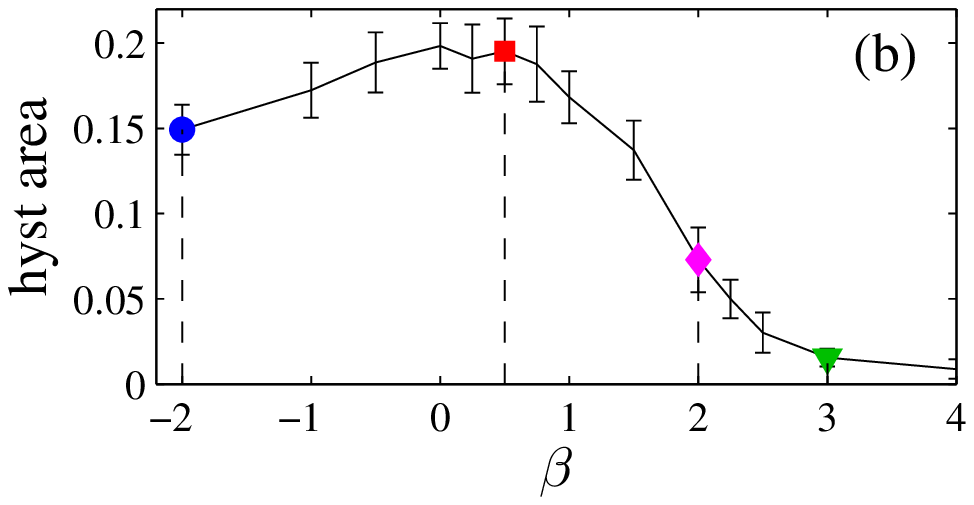}
\caption[]{(Color online). (a) Synchronization transitions for SF networks using
  different schemes of coupling weighting. In black triangles,
  the link between nodes $i$ and $j$ is weighted using $\Omega_{ij}^{\alpha}$
  with $\alpha=1$ (as in Fig. \ref{fig:hetero}), while the rest of cases refer to
  the weighting function $\widetilde\Omega_{ij}$ of Eq.(\ref{eq:weight}),
  with the values of the $\beta$ parameter given in the legend.
  In all cases, forward and backward simulations correspond respectively to continuous and dashed lines.
  (b) Area of
  the hysteretic region {\it vs.} $\beta$. Each point is an average of
  10 different simulations, each one starting from a different realization of the frequency distributions.
  In all cases, $\langle  k\rangle=30$, $N=1000$ and natural frequencies uniformly distributed in the interval $[0,1]$. }
\label{fig:betweeness}
\end{figure}

The problem closely resembles what was called, in past years, the {\it paradox of heterogeneity} \cite{nishi} where increasing the heterogeneity in the connectivity distribution of a unweighted network led to an overall deterioration
of synchrony, despite the associated reduction of the network's shortest path. That paradox was lately solved by
proving optimal synchronization conditions when proper weighting procedures are implemented on the graph's links accounting for either local \cite{motter} or global \cite{bocca} information on the specific network topology.
Therefore, in analogy with what reported in Ref. \cite{bocca}, we consider a new weighting function
\begin{equation}
 \widetilde\Omega_{ij}=a_{ij}|\omega_i-\omega_j|\displaystyle\frac{\ell_{ij}^{\beta}}{\sum_{j\in
   {\cal{N}}_i}\ell_{ij}^{\beta}},
  \label{eq:weight}
\end{equation}

\noindent
with $\beta$ being a parameter and $\ell_{ij}$ the {\it edge betweenness} associated to the link $a_{ij}$
\cite{newman}, defined as the number of shortest paths between pairs of nodes in the network that run through that
edge.

The results are reported in Fig.~\ref{fig:betweeness}(a). While the case $\beta=0$ (black triangles, already shown in Fig. 4)
corresponds to a smooth transition,  the effect for $\beta\neq 0$ in Eq.(\ref{eq:weight}) is highly nontrivial.
Precisely, moderate (positive or negative) values of $\beta$ establish in system (\ref{eq:kuramoto}) an abrupt
transition to synchronization. However, increasing $\beta$ beyond a critical value leads system (\ref{eq:kuramoto}) to display again a smooth and reversible character of the transition.

On its turn, Fig.~\ref{fig:betweeness}(b) reports the hysteresis' area
(the area of the plane ($R, \sigma/\langle k\rangle$) covered by the
hysteretic region) as a function of $\beta$, obtained by an ensemble
average over 10 different forward and backward simulations of system
(\ref{eq:kuramoto}) together with the weighting function (\ref{eq:weight}), each one starting from a different realization of the uniform frequency distribution.
The plot reveals the existence of
an {\it optimal} condition at around $\beta=0.5$ where the width of the hysteresis is maximized.
Therefore, $\beta$ can be seen as an operational parameter through which one can control and regulate the width
and extent in $\sigma$ of the hysteresis associated with the irreversible nature of ES. The latter can be of interest for controlling the range of coupling strength for which system (\ref{eq:kuramoto}) can be used to originate magnetic-like states of synchronization, i.e. situations in which an originally unsynchronized configuration, once entrained to a given phase by an external pacemaker acting for a limited time lapse, is able to permanently stay in a synchronized configuration \cite{leyvanat}.

\section{Analytical results}
In order to study the onset and nature of the explosive transition, we must analytically examine the behavior of the system in the thermodynamic limit. Let us consider the paradigmatic case in which $N$ oscillators form a fully connected graph, as the original Kuramoto model, but with weights $\Omega_{ij}=|\omega_i -\omega_j|$. Then, the dynamical equations are
\begin{equation*}
\dot{\theta}_i = \omega_i + \frac{\sigma}{N} \sum_{j=1}^N \Omega_{ij} \sin(\theta_j - \theta_i),
\end{equation*}
for $i=1,\ldots N$.

By considering the following definitions,
\begin{eqnarray*}
\frac{1}{N} \sum_{j=1}^N \Omega_{ij} \sin \theta_j &:=& A_i \sin \phi_i, \\
\frac{1}{N} \sum_{j=1}^N \Omega_{ij} \cos \theta_j &:=& A_i \cos \phi_i.
\end{eqnarray*}
the dynamical equations are usually expressed \cite{Strogatz00} in terms of
trigonometric functions  as
\begin{equation*}
\dot{\theta}_i = \omega_i + \sigma A_i \sin (\phi_i - \theta_i).
\end{equation*}

While these transformations are the same as those used in the original
Kuramoto model, now there is an explicit dependence on $i$ in 
the quantities $A_i$ and $\phi_i$. In order to continue our analysis, we will then assume some mild approximations.

In the co-rotating frame, the phases must verify  $\omega_i = \sigma A_i \sin ( \theta_i -\phi_i)$ to have a static solution (i.e., $\dot{\theta}_i=0$), which in the thermodynamic limit reads
\begin{equation}
\omega = \sigma A_\omega \sin ( \theta_\omega - \phi_\omega ). \label{eq_campomedio}
\end{equation}

The definition of $A_\omega$ and $\phi_\omega$ implies that
\begin{equation*}
F(\omega) := A_\omega \sin \phi_\omega = \int g(x) |w-x| \sin \theta (x) \, dx,
\end{equation*}
whose second derivative verifies
\begin{equation*}
F''(\omega) = \int g(x) 2 \delta(w-x) \sin \theta (x) \, dx = 2 g(\omega) \sin \theta (\omega),
\end{equation*}
using the distributional derivative of the signum function. Likewise, if we consider
\begin{equation*}
G(\omega) := A_\omega \cos \phi_\omega = \int g(x) |w-x| \cos \theta (x) \, dx,
\end{equation*}
its second derivative verifies
\begin{equation*}
G''(\omega) = 2 g(\omega) \cos \theta (\omega).
\end{equation*}
Then, Eq.~ (\ref{eq_campomedio}) takes the form
\begin{equation}
\frac{2}{\sigma} g(\omega) \omega = F''(\omega) G(\omega) - F(\omega) G''(\omega). \label{eq_derivadas}
\end{equation}

Let us work out $F(\omega)$ and $G(\omega)$. When all oscillators are close to synchronization, we can assume that $\cos \theta(x) \approx R$, thus
\begin{eqnarray*}
G(\omega) \approx R \int g(x) |w-x| \, dx = R s(\omega),
\end{eqnarray*}
where $s(\omega)$ is just the strength of a node with intrinsic frequency $\omega$. Therefore, Eq.~(\ref{eq_derivadas}) can be approximated by
\begin{equation}
\frac{2}{R \sigma} g(\omega) \omega = F''(\omega) s(\omega) - F(\omega) s''(\omega), \label{eq_ode}
\end{equation}
which is a second order ODE whose integration yields $F(\omega)$. Notice that when $s(\omega)$ is a rather involved function, Eq.~(\ref{eq_ode}) is already an approximation, and we can just consider a polynomial expansion in $\omega$ to obtain an analytical expression of $F(\omega)$.

For instance, given a uniform distribution $g(\omega)$ in the interval $[-a/2,+a/2]$, the resulting strength is a second order polynomial,
\begin{equation}
s(\omega) = a \left[ \left( \frac{\omega}{a} \right)^2 + \frac{1}{4}
\right], 
\label{fitstrength}
\end{equation}
which perfectly fits our numerical simulations (see Fig.~\ref{fig:intro}(b)), even though it has been deduced for a complete graph. Then, the integration of Eq.~(\ref{eq_ode}) results in
\begin{equation*}
F(\omega) = a \frac{ \left[ 1 + 4 \left( \frac{\omega}{a} \right)^2 \right] \arctan \left( \frac{2w}{a} \right) - (2 + \pi) \frac{w}{a}}{(4 + \pi) \sigma R},
\end{equation*}
using the initial condition $F(0)=0$, since $g(\omega)$ is a symmetric function (thus $F(\omega)$ is an odd function), and the consistency equation
\begin{equation*}
F(\omega) = \int g(x) |\omega - x| \sin \theta (x) \, dx = \int \frac{|\omega - x|}{2} F''(x) \, dx.
\end{equation*}
Therefore, since $F''(\omega) = 2 g(\omega) \sin \theta (\omega)$, we find that
\begin{equation*}
\sin \theta (\omega) = \frac{1}{\sigma R} \, H\left( \frac{2 \omega}{a} \right),
\end{equation*}
where
\begin{equation*}
H(z) := \frac{4}{4 + \pi} \left[ \frac{z}{1 + z^2} + \arctan (z) \right].
\end{equation*}

To determine how the order parameter $R$ depends on the coupling constant $\sigma$, we use that
\begin{equation}
R = \int g(x) \cos \theta (x) \, dx = \int g(x) \sqrt{1 - \sin^2 \theta (x)} \, dx, \label{eq_r}
\end{equation}
which is an implicit equation in $R$. When $\sigma R \ge \frac{2+\pi}{4+\pi} \approx 0.72$, $\sin \theta (x) \le 1$ for all $x$, which means that all oscillators are frequency locked and, then,
\begin{equation*}
R = \int_{-\frac{a}{2}}^{\frac{a}{2}} g(x) \sqrt{1- \left[ \frac{1}{\sigma R} H \left( \frac{2x}{a} \right) \right]^2} \, dx.
\end{equation*}
When $\sigma R \le \frac{2+\pi}{4+\pi}$, only those oscillators with frequency in the interval $[-\omega^*,\omega^*]$ are locked, being
\begin{equation*}
\omega^* := \frac{a}{2} H^{-1} (\sigma R),
\end{equation*}
thus
\begin{equation*}
R = \int_{-\frac{a}{2} H^{-1} (\sigma R)}^{\frac{a}{2} H^{-1} (\sigma R)} g(x) \sqrt{1- \left[ \frac{1}{\sigma R} H \left( \frac{2x}{a} \right) \right]^2} \, dx.
\end{equation*}
Hence, if we define
\begin{equation*}
\ell (\mu) := \left\lbrace
  \begin{array}{ll}
     1 & \text{ if } \mu \ge \frac{2+\pi}{4+\pi} \\
     H^{-1} (\mu) & \text{ if } 0 \le \mu < \frac{2+\pi}{4+\pi} \\
  \end{array} \right.
\end{equation*}
and
\begin{equation*}
I(\mu) := \int_{0}^{\ell(\mu)} \sqrt{1- \left[ \frac{1}{\mu} H(z) \right]^2} \, dz,
\end{equation*}
Eq.~(\ref{eq_r}) takes the form
\begin{equation}
\frac{\mu}{\sigma} = I(\mu),
\label{eq_sol}
\end{equation}
being $\mu = \sigma R$. Therefore, given a coupling constant $\sigma$, the value of $R$ is computed by solving this implicit equation in $\mu$. Notice that, geometrically, the solutions are the points where the straight line passing through the origin with slope $1/\sigma$ intersects $I(\mu)$.

\begin{figure}
  \centering
{\includegraphics[width=0.4\textwidth]{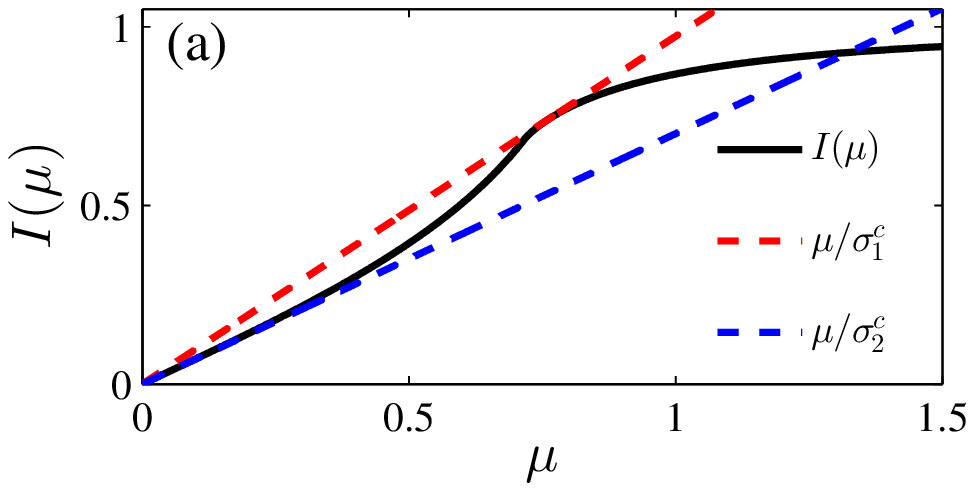}}
 \includegraphics[width=0.4\textwidth]{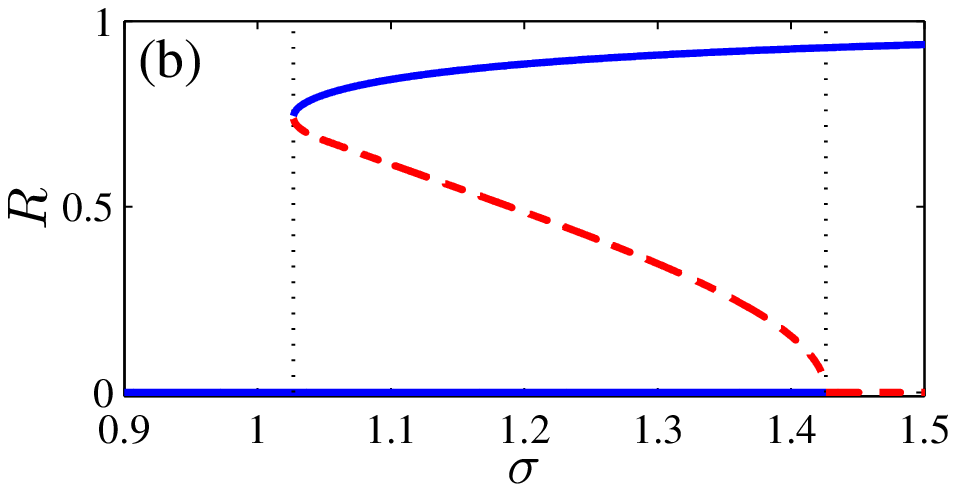}
\caption[]{(Color online). (a) $I$ as a function of $\mu=\sigma
  R$ (solid curve) as given by Eq.~(\ref{eq_sol}). The dashed lines are those straight lines whose
  intersection with $I$ marks the backward ($\sigma_1^c$) and
  forward ($\sigma_2^c$) critical points of ES for
  an all to all connected network and for a uniform frequency
  distribution. (b) The corresponding synchronization order parameter
  $R$ as a function of the coupling strength. Solid (dashed)
  curves correspond to the stable (unstable) solution. Dotted vertical
  lines mark the region of hysteresis defined by $\sigma_1^c$ and
  $\sigma_2^c$ in (a). }
\label{fig:teo}
\end{figure}

The main feature characterizing $I(\mu)$ is its inflection point at
$\frac{2+\pi}{4+\pi}$, at which the curve changes from being concave
up to concave down (see Fig.~(\ref{fig:teo})). This implies that,
depending on $\sigma$, there are three qualitatively different type of
solutions. When $\sigma$ is small, we have the trivial solution $R=0$
since the straight line and $I(\mu)$ only intersect at $\mu=0$. This
situation changes when $\sigma$ is such that the slope of the straight
line is tangent to $I(\mu)$ (i.e., when $\sigma = 1.03$, corresponding
to the red dashed line in Fig.\ref{fig:teo}(a)). When $\sigma$ is
greater than this value, we enter into the region where the hysteresis
takes place since, now, there are three values of $R$, two of them are
stable solutions ($R=0$ and $R \approx 1$) and the third one is a
unstable solution (see Fig. \ref{fig:teo}(b)). The solution $R \approx
1$ appears therefore abruptly, due to the existence of the inflection
point. This behavior changes when the slope of the straight line is
tangent to $I(0)$ (i.e., when $\sigma = 1.43$, corresponding to the
blue dashed line in Fig.\ref{fig:teo}(a)), which is the point where
the stable solution $R=0$ collapses with the unstable one, becoming
unstable (see Fig.\ref{fig:teo}(b)). Notice that the numerical values 
obtained for the Kuramoto model for large $N$ in
Fig.~\ref{fig:scaling}(b) are quite close to those predicted by the
theory.

\section{Conclusions}

In conclusion, we have introduced a weighting procedure
based on the link frequency mismatch and on the link betweenness to
induce an explosive transition to synchronization in a generic complex
network of phase oscillators and for a generic distribution of the
frequencies.  As a consequence of this procedure, topological/dynamical correlation features {\it spontaneously} emerge,
with the result of shaping a bipartite-like network where frequency
disassortativity prevails.

In this scenario, the passage from a smooth to an abrupt transition
has found to be fully rescalable, and  critically depends only on the
average connectivity, and not on the network size.

In addition, we analytically proved that our weighting procedure
yields a first-order like transition whose hysteresis extent
is calculated. Moreover, the theoretical framework allows for a geometrical
interpretation of the explosive transition in which the weighting
imposes a multi-valued Kuramoto phase order parameter, in contrast
with the classical model. 

The present results could provide significant insights into the study of
real complex networks such as power grids which can be modeled as
networks of phase oscillators whose coupling may depend on the
dynamics of the nodes \cite{motter2}.

\section*{Acknowledgments}

Authors acknowledge Alessandro Torcini for many fruitful discussion on the subject, and
the computational resources and assistance provided by CRESCO, the center of ENEA in Portici, Italy.
Financial support from the Spanish Ministerio de Ciencia e Innovaci\'on (Spain) under
projects FIS2011-25167, FIS2009-07072, and of Comunidad de Madrid (Spain) under project MODELICO-CM S2009ESP-1691, are also acknowledged.


\end{document}